\documentclass[final, a4paper]{aipproc}
\usepackage{amssymb}

\layoutstyle{8x11single}

\newcommand{\z}{\phantom{0}}
\newcommand{\ds}{\multicolumn{1}{c}{---}}

\DeclareMathAlphabet{\bm}{OT1}{ptm}{b}{it}
\begin{document}

\title[New Methods for Approximating General Relativity]
      {New Methods for Approximating General Relativity \\
       in Numerical Simulations of Stellar Core Collapse}



\author{H.~Dimmelmeier}{address={Max-Planck-Institut f\"ur
    Astrophysik, Karl-Schwarzschild-Stra{\ss}e 1, 85741 Garching,
    Germany}}

\author{P.~Cerd\'a-Dur\'an}{address={Departamento de Astronom\'\i a y
    Astrof\'\i sica, Universidad de Valencia, Dr.\ Moliner 50, 46100
    Burjassot, Valencia, Spain}}

\author{A.~Marek}{address={Max-Planck-Institut f\"ur
    Astrophysik, Karl-Schwarzschild-Stra{\ss}e 1, 85741 Garching,
    Germany}}

\author{G.~Faye}{address={Institut d'Astrophysique de Paris,
    98bis Boulevard Arago, 75014 Paris, France}}

%%%%%%%%%%%%%%%%%%%%%%%%%%%%%%%%%%%%%%%%%%%%%%%%%%%%%%%%%%%%%%%%%%%%%%%%%%%%%%%%%%%
%%%%%%%%%%%%%%%%%%%%%%%%%%%%%%%%%%%%%%%%%%%%%%%%%%%%%%%%%%%%%%%%%%%%%%%%%%%%%%%%%%%
%%%%%%%%%%%%%%%%%%%%%%%%%%%%%%%%%%%%%%%%%%%%%%%%%%%%%%%%%%%%%%%%%%%%%%%%%%%%%%%%%%%

\begin{abstract}
  We review various approaches to approximating general relativistic
  effects in hydrodynamic simulations of stellar core collapse and
  post-bounce evolution. Different formulations of a modified
  Newtonian gravitational potential are presented. Such an effective
  relativistic potential can be used in an otherwise standard
  Newtonian hydrodynamic code. An alternative approximation of general
  relativity is the assumption of conformal flatness for the
  three-metric, and its extension by adding second post-Newtonian
  order terms. Using a code which evolves the coupled system of metric
  and fluid equations, we apply the various approximation methods to
  numerically simulate axisymmetric models for the collapse of
  rotating massive stellar cores. We compare the collapse dynamics and
  gravitational wave signals (which are extracted using the quadrupole
  formula), and thereby assess the quality of the individual
  approximation method. It is shown that while the use of an effective
  relativistic potential already poses a significant improvement
  compared to a genuinely Newtonian approach, the two
  conformal-flatness-based approximation methods yield even more
  accurate results, which are qualitatively and quantitatively very
  close to those of a fully general relativistic code even for
  rotating models which almost collapse to a black hole.
\end{abstract}

\maketitle

%%%%%%%%%%%%%%%%%%%%%%%%%%%%%%%%%%%%%%%%%%%%%%%%%%%%%%%%%%%%%%%%%%%%%%%%%%%%%%%%%%%
%%%%%%%%%%%%%%%%%%%%%%%%%%%%%%%%%%%%%%%%%%%%%%%%%%%%%%%%%%%%%%%%%%%%%%%%%%%%%%%%%%%
%%%%%%%%%%%%%%%%%%%%%%%%%%%%%%%%%%%%%%%%%%%%%%%%%%%%%%%%%%%%%%%%%%%%%%%%%%%%%%%%%%%

\section{Introduction}

Gravity is the driving force which at the end of the life of massive
stars overcomes the pressure forces and causes the collapse of the
stellar core. Furthermore, the subsequent supernova explosion results
from the fact that various processes tap the enormous amount of
gravitational binding energy released during the formation of the
proto-neutron star. Therefore, gravity plays an important role during
all stages of a core collapse supernova. General relativistic effects
are important in this scenario and cannot be neglected in quantitative
models because of the strong compactness of the proto-neutron star. It
is thus essential to  include a proper treatment of general relativity
(GR) or an appropriate approximation in the numerical codes that one
uses to study core collapse supernovae.

Stellar core collapse is also among the most promising sources of
detectable gravitational waves. As the first-stage gravitational
wave interferometer detectors (like TAMA300, GEO600, LIGO, or VIRGO)
are already taking data, the interest in performing core collapse
simulations has been further motivated by the necessity of obtaining
reliable gravitational waveform predictions for detector data
analysis.

In the last two decades, a number of numerical simulations of
rotating stellar core collapse were performed, paying particular
attention to computing gravitational wave emission (see
\cite{new_03_a} and references therein). Most simulations were
done using simple analytic equations of state (EOS), and were either
restricted to Newtonian gravity or at most approximations of full
GR. Only recently there were several successful attempts to model
rotating core collapse to a proto-neutron star in full
GR~\cite{shibata_04_a, shibata_05_a}. Multi-dimensional simulations
involving a sophisticated treatment of microphysics like a tabulated
EOS and neutrino transport are up to now only performed in Newtonian
gravity (see, e.g., \cite{buras_06_a}).

In order to facilitate the numerical investigation of rotating
stellar core collapse taking into account effects of GR in ways which
are both accurate and comparatively easy to implement without having
to resort to a fully relativistic formulation, in recent years a
number of approximation scenarios were suggested. In this work we
summarize several such methods and apply them to numerical simulations
for modeling the collapse of rotating massive stellar cores. These
models, which were investigated in~\cite{shibata_05_a} in a full GR
framework, feature comparatively extreme initial masses, rotation
states, and collapse parameters, and thus reach very high densities or
even collapse to black holes. Therefore, this is an ideal test
scenario to assess the quality of the various approximation methods of
full GR.

Throughout the article, we use geometrized units with $ c = G = 1 $
and Einstein's summation convention.

%%%%%%%%%%%%%%%%%%%%%%%%%%%%%%%%%%%%%%%%%%%%%%%%%%%%%%%%%%%%%%%%%%%%%%%%%%%%%%%%%%%
%%%%%%%%%%%%%%%%%%%%%%%%%%%%%%%%%%%%%%%%%%%%%%%%%%%%%%%%%%%%%%%%%%%%%%%%%%%%%%%%%%%
%%%%%%%%%%%%%%%%%%%%%%%%%%%%%%%%%%%%%%%%%%%%%%%%%%%%%%%%%%%%%%%%%%%%%%%%%%%%%%%%%%%

\section{Hydrodynamic equations in flux-conservative hyperbolic form}

In the following we present the hydrodynamic equations in
flux-conservative formulation, which govern the evolution of matter
(approximated as an perfect fluid) in a dynamic spacetime
$ g_{\mu \nu} $. We adopt the $ 3 + 1 $ formalism to foliate
spacetime into a set of non-intersecting spacelike hypersurfaces
$ \Sigma_{\,t} $. The line element then reads
\begin{equation}
  ds^{\,2} = g_{ij} \, dx^{\,\mu} dx^{\nu} =
  - \alpha^{\,2} dt^{\,2} +
  \gamma_{\,ij} (dx^{\,i} + \beta^i dt) (dx^{\,j} + \beta^j dt),
  \label{eq:line_element}
\end{equation}
where $ \alpha $ is the lapse function describing the rate of advance
of time along a timelike unit vector $ n^{\,\mu} $ normal to
$ \Sigma_{\,t} $, $ \beta^i $ is the spacelike shift three-vector which
specifies the motion of coordinates within a surface, and
$ \gamma_{\,ij} $ is the spatial three-metric. In the Newtonian limit
$ g_{\mu \nu} $ reduces to the flat metric $ \hat{g}_{\mu \nu} $
with the associated flat three-metric $ \hat{\gamma}_{\,ij} $.

%%%%%%%%%%%%%%%%%%%%%%%%%%%%%%%%%%%%%%%%%%%%%%%%%%%%%%%%%%%%%%%%%%%%%%%%%%%%%%%%%%%
%%%%%%%%%%%%%%%%%%%%%%%%%%%%%%%%%%%%%%%%%%%%%%%%%%%%%%%%%%%%%%%%%%%%%%%%%%%%%%%%%%%

\subsection{Newtonian hydrodynamics}

As primitive (physical) hydrodynamic variables of a perfect fluid we
choose the rest-mass density $ \rho $, the covariant three-velocity
$ v_i $ as measured by an Eulerian observer at rest in $ \Sigma_{\,t} $,
and the specific internal energy $ \epsilon $. We introduce the
following set of conserved variables:
\begin{displaymath}
  D = \rho,
  \qquad
  S_i = \rho v_i,
  \qquad
  \textstyle
  \tau = \rho (\epsilon + \frac{1}{2} v_i v^{\,i}).
\end{displaymath}
Then we obtain a first-order, flux-conservative hyperbolic system of
hydrodynamic equations in Newtonian gravity,
\begin{equation}
  \frac{\partial \sqrt{\hat{\gamma}\,} \bm{U}}{\partial t} +
  \frac{\partial \sqrt{\hat{\gamma}\,} \bm{F}^{\,i}}{\partial x^{\,i}} =
  \sqrt{\hat{\gamma}\,} \bm{S},
  \label{eq:conservation_equations_n}
\end{equation}
where $ \hat{\gamma} $ is the determinant of $ \hat{\gamma}_{\,ij} $.
The state vector $ \bm{U} $, flux vector $ \bm{F}^{\,i} $, and source
vector $ \bm{S} $ are given by
\begin{displaymath}
  \bm{U} =
  [D, S_j, \tau],
  \qquad
  \bm{F}^i =
  \left[ D v^{\,i}, S_j v^{\,i} + \delta^{\,i}_j P,
  (\tau + P) v^{\,i} \right]\!,
  \qquad
  \bm{S} =
  \left[ 0, - \rho \frac{\partial \Phi}{\partial x^j},
  - \rho v^{\,i} \frac{\partial \Phi}{\partial x^{\,i}} \right]\!\!.
\end{displaymath}
Here $ \Phi $ is the Newtonian gravitational potential. An EOS, which
relates the fluid pressure $ P $ to some thermodynamically independent
quantities, e.g., $ P = P (\rho, \epsilon) $, closes the system of
conservation equations.

%%%%%%%%%%%%%%%%%%%%%%%%%%%%%%%%%%%%%%%%%%%%%%%%%%%%%%%%%%%%%%%%%%%%%%%%%%%%%%%%%%%
%%%%%%%%%%%%%%%%%%%%%%%%%%%%%%%%%%%%%%%%%%%%%%%%%%%%%%%%%%%%%%%%%%%%%%%%%%%%%%%%%%%

\subsection{General relativistic hydrodynamics}

The hydrodynamic evolution of a perfect fluid in GR with four-velocity
$ u^{\,\mu} $, rest-mass current $ J^{\,\mu} = \rho u^{\,\mu} $, and
energy-momentum tensor $ T^{\mu \nu} = \rho h u^{\,\mu} u^{\nu} + P
g^{\,\mu \nu} $ is determined by a system of local conservation
equations,
\begin{equation}
  \nabla_{\!\mu} J^{\,\mu} = 0, \qquad \nabla_{\!\mu} T^{\mu \nu} = 0,
  \label{eq:equations_of_motion_gr}
\end{equation}
where $ \nabla_{\!\mu} $ denotes the covariant derivative. The
quantity $ h = 1 + \epsilon + P / \rho $ is the specific enthalpy, and
the three-velocity is given by
$ v^{\,i} = u^{\,i} / (\alpha u^{\,0}) + \beta^i / \alpha $. Along the
same lines as in the Newtonian framework, and following the work
presented in~\cite{banyuls_97_a}, we define the set of conserved
variables as
\begin{displaymath}
  D = \rho W,
  \qquad
  S_i = \rho h W^{\,2} v_i,
  \qquad
  \tau = \rho h W^{\,2} - P - D.
\end{displaymath}
In the above expressions $ W = \alpha u^{\,0} $ is the Lorentz factor,
which satisfies the relation $ W = 1 / \sqrt{1 - \gamma^{\,ij} v_i v_{\!j}} $.

Then the local conservation laws~(\ref{eq:equations_of_motion_gr})
can again be written as a first-order, flux-conservative hyperbolic
system of hydrodynamic equations in GR gravity,
\begin{equation}
  \frac{\partial \sqrt{\gamma\,} \bm{U}}{\partial t} +
  \frac{\partial \sqrt{- g\,} \bm{F}^{\,i}}{\partial x^{\,i}} =
  \sqrt{- g\,} \bm{S},
  \label{eq:conservation_equations_gr}
\end{equation}
with
\begin{displaymath}
  \bm{U} =
  [D, S_j, \tau],
  \qquad
  \bm{F}^{\,i} =
  \left[ D \hat{v}^{\,i}\!, S_j \hat{v}^{\,i} + \delta^{\,i}_j P,
  \tau \hat{v}^{\,i} + P v^{\,i} \right]\!,
  \qquad
  \bm{S} =
  \left[ 0, T^{\mu \nu} \!\! \left( \!
  \frac{\partial g_{\nu j}}{\partial x^{\,\mu}} - 
  \Gamma^{\,\lambda}_{\!\mu \nu} g_{\lambda j} \right)\!\!,
  \alpha \left( T^{\mu 0}
  \frac{\partial \ln \alpha}{\partial x^{\,\mu}} -
  T^{\mu \nu} \Gamma^{\,0}_{\!\mu \nu} \!\!\right) \! \right]\!\!.
\end{displaymath}
Here $ \hat{v}^{\,i} = v^{\,i} - \beta^i / \alpha $, and $ g $ and
$ \gamma $ are the determinant of $ g_{ij} $ and $ \gamma_{\,ij} $,
respectively, with $ \sqrt{-g} = \alpha \sqrt{\gamma} $. In addition,
$ \Gamma^{\,\lambda}_{\!\mu \nu} $ are the Christoffel symbols
associated with $ g_{\mu \nu} $.

%%%%%%%%%%%%%%%%%%%%%%%%%%%%%%%%%%%%%%%%%%%%%%%%%%%%%%%%%%%%%%%%%%%%%%%%%%%%%%%%%%%
%%%%%%%%%%%%%%%%%%%%%%%%%%%%%%%%%%%%%%%%%%%%%%%%%%%%%%%%%%%%%%%%%%%%%%%%%%%%%%%%%%%
%%%%%%%%%%%%%%%%%%%%%%%%%%%%%%%%%%%%%%%%%%%%%%%%%%%%%%%%%%%%%%%%%%%%%%%%%%%%%%%%%%%

\section{Gravitational field equations}

%%%%%%%%%%%%%%%%%%%%%%%%%%%%%%%%%%%%%%%%%%%%%%%%%%%%%%%%%%%%%%%%%%%%%%%%%%%%%%%%%%%
%%%%%%%%%%%%%%%%%%%%%%%%%%%%%%%%%%%%%%%%%%%%%%%%%%%%%%%%%%%%%%%%%%%%%%%%%%%%%%%%%%%

\subsection{Effective relativistic potential for Newtonian simulations}

%%%%%%%%%%%%%%%%%%%%%%%%%%%%%%%%%%%%%%%%%%%%%%%%%%%%%%%%%%%%%%%%%%%%%%%%%%%%%%%%%%%

\noindent
{\bf \emph{Method~N: Newtonian potential for a self-gravitating
    fluid.\quad}}
The Newtonian potential $ \Phi $, whose gradient acts as an external
force in the source of the Newtonian hydrodynamic
equations~(\ref{eq:conservation_equations_n}), is determined by the
Poisson equation
\begin{equation}
  \Delta \Phi = 4 \pi \rho.
  \label{eq:newtonian_potential}
\end{equation}
To approximately include effects of GR gravity, several successful
attempts of increasing complexity have been made~\cite{rampp_02_a,
  marek_06_a, mueller_06_a} to replace the regular Newtonian potential
$ \Phi $ (called \emph{method~N} here), by an effective relativistic
gravitational potential $ \Phi_\mathrm{eff} $ for a self-gravitating
fluid, which mimics the deeper gravitational well of the GR
case. \\

%%%%%%%%%%%%%%%%%%%%%%%%%%%%%%%%%%%%%%%%%%%%%%%%%%%%%%%%%%%%%%%%%%%%%%%%%%%%%%%%%%%

\noindent
{\bf \emph{Method~R: TOV potential for a self-gravitating
    fluid.\quad}}
As a first approximation, we demand that $ \Phi_\mathrm{eff} $
in spherical symmetry reproduces the solution of hydrostatic
equilibrium according to the Tolman--Oppenheimer--Volkoff (TOV)
equation. To obtain an effective relativistic potential also for
a nonspherical matter distribution, we construct a generalized TOV
potential~\cite{rampp_02_a} as a radial integral over angular averaged
matter quantities $ \overline{\!\rho} $, $ \overline{\!\epsilon} $,
$ \overline{v_r} $, and $ \overline{\!P} $:
\begin{equation}
  \overline{\!\Phi\!}\,_\mathrm{\,TOV} (r) =
  - \int_r^{\,\infty} \frac{dr\,'}{r\,'^{\,2}}
  \left( \overline{\!m} + 4 \pi r\,'^{\,3} \overline{\!P} \right)
  \frac{\overline{\!h}}{\overline{\Gamma}^{\,2}}.
  \label{eq:tov_potential}
\end{equation}
Here $ \overline{v_r} $ is the local radial velocity of the fluid.
The TOV mass $ \overline{\!m} $ and metric function
$ \overline{\Gamma} $ are given by
\begin{equation}
  \overline{\!m} (r) = 4 \pi \int_0^{\,r} dr\,' \,
  r\,'^{\,2} \, \overline{\!\rho} \left( 1 + \overline{\!\epsilon} \right),
  \qquad
  \overline{\Gamma} =
  \sqrt{1 + \overline{v_r}^{\,2} - \frac{2 \overline{\!m}}{r}}.
  \label{eq:tov_mass_and_gamma_factor}
\end{equation}

Then the effective relativistic potential $ \Phi_\mathrm{eff} $ can be
calculated by substituting the ``spherical contribution''
\begin{equation}
  \overline{\!\Phi\!}\, (r) = - 4 \pi \int_0^{\,\infty} \!\!
  \mathrm{d}r\,' \, r\,'^{\,2} \, \frac{\overline{\!\rho}}{|r - r\,'|}
  \label{eq:spherical_newtonian_potential}
\end{equation}
to the multi-dimensional Newtonian gravitational potential $ \Phi $
by the TOV potential $ \overline{\!\Phi\!}\,_\mathrm{\,TOV} $:
\begin{equation}
  \Phi_\mathrm{eff} = \Phi - \overline{\!\Phi\!}\, +
  \overline{\!\Phi\!}\,_\mathrm{\,TOV}.
  \label{eq:effective_tov_potential}
\end{equation}

When this reference case of the effective relativistic potential,
\emph{method~R}, is used in an otherwise Newtonian simulation, the
collapse of a spherical stellar core to a proto-neutron star yields
after core bounce and ring-down a new quasi-equilibrium state which is
the exact (numerical) solution of the TOV structure equations of GR.

The effective relativistic potential of \emph{method~R} has already
been used in simulations of supernova core collapse in spherical
symmetry with sophisticated microphysics including Boltzmann neutrino
transport~\cite{rampp_02_a, liebendoerfer_05_a} as well as for
rotating stellar core collapse with a simple matter treatment without
magnetic fields~\cite{marek_06_a} and with magnetic
fields~\cite{obergaulinger_06_a}. \\

%%%%%%%%%%%%%%%%%%%%%%%%%%%%%%%%%%%%%%%%%%%%%%%%%%%%%%%%%%%%%%%%%%%%%%%%%%%%%%%%%%%

\noindent
{\bf \emph{Method~A: Modifications of the TOV potential.\quad}}
In a recent comparison~\cite{liebendoerfer_05_a} of supernova core
collapse in spherical symmetry it was found that the use of the
effective relativistic potential in the
form~(\ref{eq:effective_tov_potential}) overrates the relativistic
effects, because in combination with Newtonian kinematics it tends to
overestimate the infall velocities and to underestimate the flow
inertia in the preshock region. As a consequence, the compactness of
the proto-neutron star is overestimated.

In order to reduce these discrepancies, several modifications of the
spherical TOV potential~(\ref{eq:tov_potential}) have been tested
recently~\cite{marek_06_a}. In all cases the construction of the
multi-dimensional effective relativistic potential
$ \Phi_\mathrm{eff} $ according to
Eq.~(\ref{eq:effective_tov_potential}) remains unchanged. In the most
accurate of the variations introduced in~\cite{marek_06_a},
\emph{method~A}, an additional factor $ \overline{\Gamma} $ is
added in the integrand of the equation for the TOV
mass~(\ref{eq:tov_mass_and_gamma_factor}). Since
$ \overline{\Gamma} < 1 $ this reduces the gravitational TOV mass used
in the TOV potential~(\ref{eq:tov_potential}), and thus attenuates the
relativistic corrections in the effective relativistic potential
$ \Phi_\mathrm{eff} $.

As a consequence, a Newtonian simulation of core collapse using the
effective relativistic potential of \emph{method~A} not only
reproduces the solution of the TOV structure equations for a matter
configuration in equilibrium to fair accuracy, but contrary to
\emph{method~R} also closely matches the results from a relativistic
simulation during the dynamic phase of the contraction of the
core. This has been demonstrated in numerical studies of collapsing
stellar cores in spherical symmetry with sophisticated microphysics
including Boltzmann neutrino transport~\cite{marek_06_a}.

The results presented in that work also demonstrate that in
axisymmetric collapse calculations for rotating stellar cores with a
wide range of initial conditions and a simple EOS, the new effective
relativistic potential of \emph{method~A} reproduces the
characteristics of the relativistic collapse dynamics quantitatively
much better than the Newtonian potential for not too rapid rotation.

%%%%%%%%%%%%%%%%%%%%%%%%%%%%%%%%%%%%%%%%%%%%%%%%%%%%%%%%%%%%%%%%%%%%%%%%%%%%%%%%%%%
%%%%%%%%%%%%%%%%%%%%%%%%%%%%%%%%%%%%%%%%%%%%%%%%%%%%%%%%%%%%%%%%%%%%%%%%%%%%%%%%%%%

\subsection{Conformal-flatness-based approximations for general
  relativistic simulations}

In the $ 3 + 1 $ formalism, the Einstein equations split into a
coupled set of 12 evolution equations for the three-metric
$ \gamma_{\,ij} $ as well as the extrinsic curvature $ K_{\,ij} $, and
4 constraint equations. We use the ADM gauge, where $ \gamma_{\,ij} $
can be decomposed into a conformally flat term with conformal factor
$ \phi $ plus a transverse traceless term:
\begin{equation}
  \gamma_{\,ij} = \phi^4 \hat{\gamma}_{\,ij} + h^\mathrm{TT}_{\,ij},
  \qquad \qquad \mbox{with} \qquad
  h^\mathrm{TT}_{\,ij} \hat{\gamma}^{\;ij} = 0,
  \qquad
  \hat{\gamma}^{\;ik} \hat{\nabla}_{\!k\,} h^\mathrm{TT}_{\,ij} = 0,
  \label{eq:adm_gauge}
\end{equation}

%%%%%%%%%%%%%%%%%%%%%%%%%%%%%%%%%%%%%%%%%%%%%%%%%%%%%%%%%%%%%%%%%%%%%%%%%%%%%%%%%%%

\noindent
{\bf \emph{Method~CFC: Conformal flatness condition for the
    three-metric.\quad}}
In spherical symmetry $ h^\mathrm{TT}_{\,ij} = 0 $, i.e.\ the
three-metric is conformally flat. Thus a reasonable approximation of
$ \gamma_{\,ij} $ for scenarios which do not deviate to strongly from
sphericity is to impose a vanishing $ h^\mathrm{TT}_{\,ij} $ in
Eq.~(\ref{eq:adm_gauge}):
\begin{equation}
  \gamma_{\,ij} = \phi^4 \hat{\gamma}_{\,ij}.
  \label{eq:cfc_metric}
\end{equation}
This is the conformal flatness condition (CFC) or
Isenberg--Wilson--Mathews approximation~\cite{isenberg_78_a,
  wilson_96_a}. Using CFC and assuming the maximal slicing condition,
for which the trace of the extrinsic curvature vanishes, the $ 3 + 1 $
metric equations reduce to a set of five coupled elliptic
(Poisson-like) equations for the metric components,
\begin{eqnarray}
  \hat{\Delta} \phi & = & - 2 \pi \phi^5
  \left( \rho h W^{\,2} - P +
  \frac{K_{\,ij} K^{\,ij}}{16 \pi} \right)\!,
  \label{eq:cfc_metric_equation_1} \\
  \hat{\Delta} (\alpha \phi) & = & 2 \pi \alpha \phi^5
  \left( \rho h (3 W^{\,2} - 2) + 5 P +
  \frac{7 K_{\,ij} K^{\,ij}}{16 \pi} \right)\!,
  \label{eq:cfc_metric_equation_2} \\
  \hat{\Delta} \beta^i & = &
  16 \pi \alpha \phi^4 S^{\,i} + 2 \phi^{10} K^{\,ij}
  \hat{\nabla}_{\!\!j} \! \left( \! \frac{\alpha}{\phi^6} \! \right) \! -
  \frac{1}{3} \hat{\nabla}^{i\,} \hat{\nabla}_{\!k\,} \beta^k\!,
  \label{eq:cfc_metric_equation_3}
\end{eqnarray}%
where $ \hat{\nabla} $ and $ \hat{\Delta} =
\hat{\gamma}^{\;ij} \hat{\nabla}_{i\,} \hat{\nabla}_{\!\!j} $
are the flat space Nabla and Laplace operators. Additionally, the
expression for the extrinsic curvature becomes time-independent and
reads $ K_{\,ij} = \frac {1}{2 \alpha} (\nabla_{\!i\,} \beta_j +
\nabla_{\!\!j\,} \beta_i - \frac{2}{3} \gamma_{\,ij} \nabla_{\!k\,}
\beta^k) $. As the CFC metric
equations~(\ref{eq:cfc_metric_equation_1}\,--\,\ref{eq:cfc_metric_equation_3})
do not contain explicit time derivatives, the metric is calculated by
a fully constrained approach.

The accuracy of the CFC approximation has been tested in various
works, e.g., for stellar core collapse and equilibrium models of
neutron stars~\cite{shibata_04_a, cook_96_a, dimmelmeier_02_a,
  dimmelmeier_05_a, saijo_05_a, dimmelmeier_06_a}, as well as for
binary neutron star merger~\cite{oechslin_02_a, faber_04_a}. The
spacetime of rapidly (uniformly or differentially) rotating neutron
star models is very well approximated by the CFC
metric~(\ref{eq:cfc_metric}). Its accuracy degrades only in extreme
cases, such as rapidly rotating massive neutron stars or black
holes. \\

%%%%%%%%%%%%%%%%%%%%%%%%%%%%%%%%%%%%%%%%%%%%%%%%%%%%%%%%%%%%%%%%%%%%%%%%%%%%%%%%%%%

\noindent
{\bf \emph{Method~CFC+: Inclusion of the second post-Newtonian deviation
  from isotropy.\quad}}
To improve the accuracy of CFC, we have developed the CFC+
approximation. It consists in adding to the CFC
metric~(\ref{eq:cfc_metric}) the second post-Newtonian (2PN) deviation
from isotropy,
\begin{equation}
  \gamma_{\,ij} = \phi^4 \hat{\gamma}_{\,ij} + h^\mathrm{\,2PN}_{\,ij}
  \label{eq:cfc_plus_metric}
\end{equation}
which includes a traceless and transverse term
$ h^\mathrm{\,2PN}_{\,ij} $. In the decomposition~(\ref{eq:adm_gauge})
$ h^\mathrm{\,2PN}_{\,ij} $ is identical to $ h^\mathrm{TT}_{\,ij} $
up to 2PN order. As described in detail in~\cite{cerda_05_a} this
correction is the solution of the tensor Poisson
equation~\cite{schaefer_90_a}
\begin{equation}
  \hat{\Delta} h^\mathrm{\,2PN}_{\,ij} =
  \hat{\gamma}_{\,ij}^{\mathrm{\,TT}kl}
  \left( \! - 4 \hat{\nabla}_{\!k\,} U \, \hat{\nabla}_{\!l\,} U -
  16 \pi \left( S^*_k S^*_l \right) /D^* \! \right)\!,
  \label{eq:cfc_plus_correction}
\end{equation}
where $ U $ is a Newtonian-like potential which is the solution of the
Poisson equation $ \hat{\Delta} U = - 4 \pi D^* $,
$ D^* = \sqrt{\gamma / \hat{\gamma}\,} D $, and
$ S^{*}_i = \sqrt{\gamma / \hat{\gamma}\,} S_i $. The
integro-differential operator $ \hat{\gamma}_{\,ij}^{\mathrm{TT}kl} $
is the traceless and transverse projector defined in the Appendix
of~\cite{cerda_05_a}. Of the CFC metric
equations~(\ref{eq:cfc_metric_equation_1}\,--\,\ref{eq:cfc_metric_equation_3}),
only the equation~(\ref{eq:cfc_metric_equation_2}) for the lapse
function $ \alpha $ is affected by the corrections
$ h^\mathrm{\,2PN}_{\,ij} $:
\begin{equation}
  \hat{\Delta} (\alpha \phi) =
  \left[ \hat{\Delta} \left( \alpha \phi \right)
  \right]_{\!h_{\,ij}^\mathrm{\,2PN} = 0} \! - 
  \hat{\gamma}^{\,ik} \hat{\gamma}^{\,jl} \,
  h^\mathrm{\,2PN}_{\,ij} \, \hat{\nabla}_{\!k\,} \hat{\nabla}_{\!l\,} U.
  \label{eq:cfc_plus_metric_equation}
\end{equation}
However, the CFC+ corrections couple implicitly to the other metric
equations as their source terms depend on $ \alpha $.

Calculating $ h^\mathrm{\,2PN}_{\,ij} $ by inversion of the tensor
Poisson equation~(\ref{eq:cfc_plus_correction}) is simplified
considerably by introducing potentials that are solutions of
scalar/vector/tensor Poisson equations~\cite{cerda_05_a}.

Applying \emph{method~CFC+} to axisymmetric simulations of pulsations
in rotating neutron stars and rotational stellar core collapse to a
proto-neutron star, it was recently found that relevant quantities
show only minute differences with respect to
\emph{method~CFC}~\cite{cerda_05_a}. By contrast, in scenarios where
the CFC approximation is expected to become increasingly inaccurate,
CFC+ is expected to yield results which are closer to full GR.

%%%%%%%%%%%%%%%%%%%%%%%%%%%%%%%%%%%%%%%%%%%%%%%%%%%%%%%%%%%%%%%%%%%%%%%%%%%%%%%%%%%
%%%%%%%%%%%%%%%%%%%%%%%%%%%%%%%%%%%%%%%%%%%%%%%%%%%%%%%%%%%%%%%%%%%%%%%%%%%%%%%%%%%
%%%%%%%%%%%%%%%%%%%%%%%%%%%%%%%%%%%%%%%%%%%%%%%%%%%%%%%%%%%%%%%%%%%%%%%%%%%%%%%%%%%

\section{Simulations of the collapse of rotating massive stellar cores}

%%%%%%%%%%%%%%%%%%%%%%%%%%%%%%%%%%%%%%%%%%%%%%%%%%%%%%%%%%%%%%%%%%%%%%%%%%%%%%%%%%%
%%%%%%%%%%%%%%%%%%%%%%%%%%%%%%%%%%%%%%%%%%%%%%%%%%%%%%%%%%%%%%%%%%%%%%%%%%%%%%%%%%%

\subsection{Setup of models and numerical methods}

We follow the work in~\cite{shibata_05_a} to set up rapidly rotating
initial models of massive stellar cores with baryonic rest masses
$ M_\mathrm{\,r} \gtrsim 1.5 \, M_\odot $, supported by electron
degeneracy pressure and a possible radiation pressure
contribution. Using Hachisu's self-consistent field
method~\cite{komatsu_89_a}, we construct the initial model as a
rotating perfect fluid in equilibrium obeying a polytropic EOS,
$ P = K_{\,0} \rho^{\gamma_{\,0}} $ and
$ \epsilon = P / [\rho (\gamma_{\,0} - 1)] $, where
$ \gamma_{\,0} = 4 / 3 $ is the adiabatic index and the polytropic
constant assumes values $ K_{\,0} = \{5.0,~7.0,~8.0\} \times 10^{14} $
(in cgs units) with increasing contribution from radiation
pressure. The initial core's central density is
$ \rho_\mathrm{\,c\;0} = 10^{10} \mathrm{\ g\ cm}^{-3} $. The rotation
law is set by $ j = A^2 (\Omega_\mathrm{\,c} - \Omega) $, where $ j $
is the specific angular momentum, $ \Omega_\mathrm{\,c} $ is the value
of the angular velocity $ \Omega $ at the center, and the constant
$ A $ determines the degree of differential rotation. We parameterize
the rotational state by two quantities, the rotation rate
$ \beta_{\,0} $ which is the ratio of rotational kinetic energy to
gravitational binding energy, and the parameter
$ \hat{A} = A / r_\mathrm{\,e} = \{\infty,~0.25,~0.1\} $, where
$ r_\mathrm{\,e} $ is the stellar equatorial radius, which decreases
as the differentiality of the rotation profile increases (with
$ \hat{A} = \infty $ corresponding to uniform rotation). To initiate
the collapse we reduce the pressure,
$ P = K_{\,0} \rho^{\gamma_{\,0}} (\gamma_{\,1} - 1) / (\gamma_{\,0} - 1) $,
with lowering the adiabatic index to
$ \gamma_{\,1} = 1.31 < \gamma_{\,0} $ and setting the polytropic
constant to $ K_{\,1} = 5.0 \times 10^{14} $. Note that this procedure
is different from the collapse initiation for the models presented
in~\cite{shibata_04_a, dimmelmeier_02_a, dimmelmeier_05_a, cerda_05_a,
  dimmelmeier_02_b}, where both the pressure and the internal energy
are reduced.

As in~\cite{shibata_04_a, shibata_05_a, dimmelmeier_02_a,
  dimmelmeier_05_a, dimmelmeier_02_b}, during the collapse phase we
use a simple hybrid ideal gas EOS~\cite{janka_93_a} that consists of a
polytropic contribution $ P_\mathrm{\,p} $ describing the degenerate
electron pressure and at supranuclear densities the pressure due to
repulsive nuclear forces, and a thermal contribution
$ P_\mathrm{\,th} $ which accounts for the heating of the matter by
shocks, $ P = P_\mathrm{\,p} + P_\mathrm{\,th} $, where
$ P_\mathrm{\,p} = K_{\,1,2} \, \rho^{\gamma_{\,1,2}} $ and
$ P_\mathrm{\,th} = \rho \epsilon_\mathrm{\,th} (\gamma_\mathrm{\,th} - 1) $.
To approximate the stiffening of the EOS at densities larger than
nuclear matter density
$ \rho_\mathrm{\,nuc} = 2.0 \times 10^{14} \mathrm{\ g\ cm}^{-3} $, the
adiabatic index increases to $ \gamma_{\,2} = 2.5 > \gamma_{\,1} $.
For the adiabatic index of the thermal contribution we choose
$ \gamma_\mathrm{\,th} = 1.3 $. The thermal internal energy is
calculated as
$ \epsilon_\mathrm{\,th} = \epsilon - \epsilon_\mathrm{\,p} $,
while the polytropic specific internal energy
$ \epsilon_\mathrm{\,p} $ is determined from $ P_\mathrm{\,p} $ by the
ideal gas relation in combination with continuity conditions, which
also determine the value for the polytropic constant $ K_{\,2} $ at
$ \rho > \rho_\mathrm{\,nuc} $ (for more details,
see~\cite{dimmelmeier_02_a, janka_93_a}). Of the 27 axisymmetric
collapse models of rotating massive stellar cores investigated
in~\cite{shibata_05_a}, here we select 8 representative
models. Table~\ref{tab:model_summary} summarizes their parameters and
establishes the nomenclature of model names.

\begin{table}[!b]
  \begin{tabular}{l|c|c|c|c|c|c|c|c}
    \hline
    Model &
    M5a1 &
    M5c2 &
    M7a4 &
    M7b1 &
    M7c3 &
    M8a1 &
    M8c2 &
    M8c4 \\
    \hline
    $ K_{\,0}~[10^{14}] $ &
    5.0\z &
    5.0\z &
    7.0\z &
    7.0\z &
    7.0\z &
    8.0\z &
    8.0\z &
    8.0\z \\
    $ \hat{A} $ &
    $ \infty $ &
    0.10 &
    $ \infty $ &
    0.25 &
    0.10 &
    $ \infty $ &
    0.25 &
    0.10 \\
    $ \beta_{\,0}~[\%] $ &
    0.89 &
    1.24 &
    3.67 &
    2.18 &
    1.27 &
    0.88 &
    2.19 &
    1.27 \\    
    \hline
  \end{tabular}
  \caption{Summary of the model parameters for rotating massive
    stellar cores.}
  \label{tab:model_summary}
\end{table}

Our evolution code, which was introduced in~\cite{dimmelmeier_02_a,
  cerda_05_a}, utilizes Eulerian spherical polar coordinates
$ \{r, \theta\} $ restricted to axisymmetry. We also assume symmetry
with respect to the equatorial plane. The finite difference grid
consists of $ n_r = 250 $ logarithmically spaced radial and
$ n_\theta = 45 $ equidistantly spaced angular grid points with a
central resolution $ \Delta r_\mathrm{\,c} \simeq 250 \mathrm{\ m} $.
A small part of the grid covers an artificial low-density atmosphere
extending beyond the stellar surface.

The hydrodynamic solver performs the time integration of
the system of conservation equations, given either by
Eq.~(\ref{eq:conservation_equations_n}) for \emph{methods~N},
\emph{R}, and~\emph{A}, or by Eq.~(\ref{eq:conservation_equations_gr})
for \emph{methods~CFC} and~\emph{CFC+}. This is done by means of a
high-resolution shock-capturing (HRSC) scheme with third-order
accurate PPM reconstruction (for a review of such methods in numerical
GR, see~\cite{font_03_a}). The numerical fluxes are computed by
Marquina's approximate flux formula~\cite{donat_98_a}. The time update
of the state vector $ \bm{U} $ is done using the method of lines in
combination with a Runge--Kutta scheme with second order accuracy in
time. Once $ \bm{U} $ is updated in time, the primitive variables are
recovered either directly (for \emph{methods~N}, \emph{R},
and~\emph{A}) or through an iterative Newton--Raphson method (for
\emph{methods~CFC} and~\emph{CFC+}).

When using \emph{methods~N}, \emph{R}, or~\emph{A} we solve the
multi-dimensional linear Poisson
equation~(\ref{eq:newtonian_potential}) for the Newtonian potential
$ \Phi $ by transforming it into an integral equation using a
Green's function. The volume integral is numerically evaluated by
expanding the source term into a series of radial functions and
associated Legendre polynomials, which we cut at order $ l = 10 $. It
is straightforward to solve the integral
equations~(\ref{eq:tov_potential}\,--\,\ref{eq:spherical_newtonian_potential})
of the relativistic corrections needed to finally obtain the effective
relativistic potential $ \Phi_\mathrm{eff} $.

The metric solver for obtaining the CFC and CFC+ spacetime
metric used in the GR simulations exploits the fact that
Eqs.~(\ref{eq:cfc_metric_equation_1}\,--\,\ref{eq:cfc_metric_equation_3}),
supplemented by Eqs.~(\ref{eq:cfc_plus_correction},
\ref{eq:cfc_plus_metric_equation}) in the case of \emph{method~CFC+},
are written in the form of a system of nonlinear coupled equations for
a vector of unknowns $ \bm{u} $ with a Laplace operator on the
left hand side, $ \hat{\Delta} \, \bm{u} = \bm{S} (\bm{u}) $. A common
method to solve such equations is to keep $ \bm{S} (\bm{u}) $ fixed,
start with an initial guess $ \bm{u}^{\,0} $, and then solve each of
the resulting decoupled linear Poisson equations
$ \hat{\Delta} \, \bm{u}^{\,s + 1} = \bm{S} (\bm{u}^{\,s}) $ in an
iteration cycle with steps $ s $ until convergence. To obtain the
solution of the now uncoupled, linear Poisson equations, we again
employ the expansion described above. More details are presented in
the description of Solver~2 in~\cite{dimmelmeier_05_a}.

The calculation of the CFC or CFC+ metric is computationally very
expensive. Hence, for these methods the metric is updated only once
every 100/10/50 hydrodynamic time steps before/during/after core
bounce, and extrapolated in between, as described
in~\cite{dimmelmeier_02_a}. We also note that we have performed grid
resolution tests to ascertain that the grid resolution specified above
is appropriate. When using the CFC/CFC+ metric equations, the degrees
of freedom representing gravitational waves are removed from the
spacetime. Therefore, in all simulations gravitational waveforms are
obtained in a post-processing step with the help of the quadrupole
formula.

%%%%%%%%%%%%%%%%%%%%%%%%%%%%%%%%%%%%%%%%%%%%%%%%%%%%%%%%%%%%%%%%%%%%%%%%%%%%%%%%%%%
%%%%%%%%%%%%%%%%%%%%%%%%%%%%%%%%%%%%%%%%%%%%%%%%%%%%%%%%%%%%%%%%%%%%%%%%%%%%%%%%%%%

\subsection{Results and discussion}

Comparing typical quantities that describe the collapse dynamics
(like the maximum density $ \rho_\mathrm{\,max\;b} $, maximum rotation
rate $ \beta_\mathrm{\,max\;b} $, or minimal value of the lapse
$ \alpha_\mathrm{\,min\;b} $ during core bounce) from our approximate
GR simulations with those obtained in full GR~\cite{shibata_05_a}, we
find excellent agreement for both \emph{method~CFC} and~\emph{CFC+}.
In the case of models which do not collapse to a black hole, our
results for $ \rho_\mathrm{\,max\;b} $ (which we present in
Table~\ref{tab:method_comparison_density}) deviate from the full GR
ones by typically less than about 2\% for models with moderate core
mass and moderate or rapid rotation. The deviation for
$ \rho_\mathrm{\,max\;b} $ increases to at most 20\% for the most
extreme model~M8c4, which has a baryonic rest mass of
$ M_\mathrm{\,r} = 3.05 $ and a slow rotation rate. Consequently,
this model reaches $ \alpha_\mathrm{\,min} = 0.29 $ and thus almost
collapses to a black hole.

\begin{table}[!b]
  \begin{tabular}{l|lc|lc|lc|lc|lc|l}
    \hline
    Model &
    \multicolumn{2}{c}{\emph{Method~N}} &
    \multicolumn{2}{|c}{\emph{Method~R}} &
    \multicolumn{2}{|c}{\emph{Method~A}} &
    \multicolumn{2}{|c}{\emph{Method~CFC}} &
    \multicolumn{2}{|c}{\emph{Method~CFC+}} &
    \multicolumn{1}{|c}{GR} \\
    \hline
    M5a1 & \z3.8 & (0.58) & \z6.1 & (0.92) &\z5.6 & (0.85) &
    \z6.6 & (1.00) & \z6.6 & (1.00) & \z6.6 \\
    M5c2 & \z1.1 & (0.22) & \z3.3 & (0.66) & \z2.9 & (0.58) &
    \z4.9 & (0.98) & \z4.9 & (0.98) & \z5.0 \\
    M7a4 & \z5.6 & --- & \ds & --- & 14 & --- &
    14 & --- & 14 & --- & \ds \\
    M7b1 & \z0.10 & (0.13) & \z0.40 & (0.51) & \z0.31 & (0.39) &
    \z0.83 & (1.05) & \z0.85 & (1.08) & \z0.79 \\
    M7c3 & \z1.2 & (0.13) & \z5.3 & (0.58) & \z4.2 & (0.46) &
    \z9.2 & (1.00) & \z9.2 & (1.00) & \z9.2 \\
    M8a1 & \z4.5 & --- & \ds & --- & 17 & --- &
    \ds & --- & \ds & --- & \ds \\
    M8c2 & \z0.19 & (0.04) & \z1.5 & (0.28) & \z9.0 & (0.17) &
    \z5.3 & (0.98) & \z5.2 & (0.96) & \z5.4 \\
    M8c4 & \z1.2 & (0.08) & \z7.1 & (0.47) & \z5.1 & (0.34) &
    17 & (1.13) & 12 & (0.80) & 15 \\
    \hline
  \end{tabular}
  \caption{Maximum density $ \rho_\mathrm{\,max\;b} $ in units of
    $ 10^{14} \mathrm{\ g\ cm}^{-3} $ during core bounce for the
    investigated rotating core collapse models from simulations
    using \emph{methods~N}, \emph{R}, \emph{A}, \emph{CFC},
    and~\emph{CFC+}. The relative deviations
    $ \rho_\mathrm{\,max\;b} / \rho_\mathrm{\,max\;b}^\mathrm{GR} $
    from the value obtained in a full GR simulation (given in
    the last column) are shown in parentheses. The dash denotes
    models for which the approximate or full GR simulation produces
    a black hole.}
  \label{tab:method_comparison_density}
\end{table}

When the less accurate approximation \emph{methods~A} and~\emph{R} are
used, the values for $ \rho_\mathrm{\,max\;b} $ as expected deviate
more strongly from full GR and approach the ones obtained using a
genuinely Newtonian simulation. Concerning the quality of our
approximations, we find equivalent results for the values of
$ \rho_\mathrm{\,max\;b} $, $ \beta_\mathrm{\,max\;b} $, and
$ \alpha_{\,min\;b} $ (with the latter obviously only defined for
\emph{methods~CFC} and~\emph{CFC+}). Note also that though
\emph{method~A} compared to \emph{method~R} yields results closer to
full GR in spherical symmetry, the influence of rotation reverses this
behavior, as already discussed in~\cite{marek_06_a}.

A very important property of any approximation method for GR
simulations is that it should correctly reproduce the
collapse type of a specific model for a range in parameter space as
broad as possible. For the investigated models,
following~\cite{shibata_05_a} we denote by BH, NS, O-A, and O-B the
collapse to a black hole, a neutron star, an oscillating star with 
$ \rho_\mathrm{\,max\;b} \ge \rho_\mathrm{\,nuc} $, and an oscillating
star with $ \rho_\mathrm{\,max\;b} < \rho_\mathrm{\,nuc} $ as end
state, respectively. As shown in
Table~\ref{tab:method_comparison_collapse_type}, \emph{methods~CFC}
and~\emph{CFC+} correctly capture the collapse type for all
investigated cases. The identification of the end state of model~M7a4
(as well as of model~M7a3) as a black hole in~\cite{shibata_05_a} is
most probably due to insufficient resolution of the inner parts of the
numerical grid, which results in fall-back to the already formed
proto-neutron star leading to its recollapse to a black hole. We find
that behavior in a resolution study of these two models. This is also
supported by independent simulations in full GR without symmetry
assumptions~\cite{shibata_05_b}. We thus classify model~M7a4 as NS.

\begin{table}[!b]
  \begin{tabular}{l|c|c|c|c|c|c}
    \hline
    Model &
    \emph{Method~N} &
    \emph{Method~R} &
    \emph{Method~A} &
    \emph{Method~CFC} &
    \emph{Method~CFC+} &
    GR \\
    \hline
    M5a1 & \framebox{NS} & \framebox{NS} & \framebox{NS} &
    \framebox{NS} & \framebox{NS} & \framebox{NS} \\
    M5c2 & O-B & O-A & O-A & \framebox{O-A $ \rightarrow $ NS} &
    \framebox{O-A $ \rightarrow $ NS} &
    \framebox{O-A $ \rightarrow $ NS} \\
    M7a4 & \framebox{NS} & BH & \framebox{NS} & \framebox{NS} &
    \framebox{NS} & \framebox{NS} / BH \\
    M7b1 & \framebox{O-B} & \framebox{O-B} & \framebox{O-B} &
    \framebox{O-B} & \framebox{O-B} & \framebox{O-B} \\
    M7c3 & O-B & \framebox{NS} & O-A $ \rightarrow $ NS &
    \framebox{NS} & \framebox{NS} & \framebox{NS} \\
    M8a1 & NS & \framebox{BH} & NS & \framebox{BH} & \framebox{BH} &
    \framebox{BH} \\
    M8c2 & O-B & O-B & O-B & \framebox{O-A} & \framebox{O-A} &
    \framebox{O-A} \\
    M8c4 & O-B & \framebox{NS} & \framebox{NS} & \framebox{NS} &
    \framebox{NS} & \framebox{NS} \\
    \hline
  \end{tabular}
  \caption{Collapse type of the investigated rotating core collapse
    models from simulations using \emph{methods~N}, \emph{R},
    \emph{A}, \emph{CFC}, and~\emph{CFC+}. The simulations which yield
    the correct collapse type obtained in a full GR simulation (given
    in the last column) are marked by a box. See the main text for the
    definition of the collapse types. For model~M7a4 the full GR
    simulation in~\cite{shibata_05_a} predicts BH instead of NS as
    final state, which is apparently due to lack of resolution.}
  \label{tab:method_comparison_collapse_type}
\end{table}

\begin{figure}[!b]
  \resizebox{1.0\columnwidth}{!}
  {\includegraphics{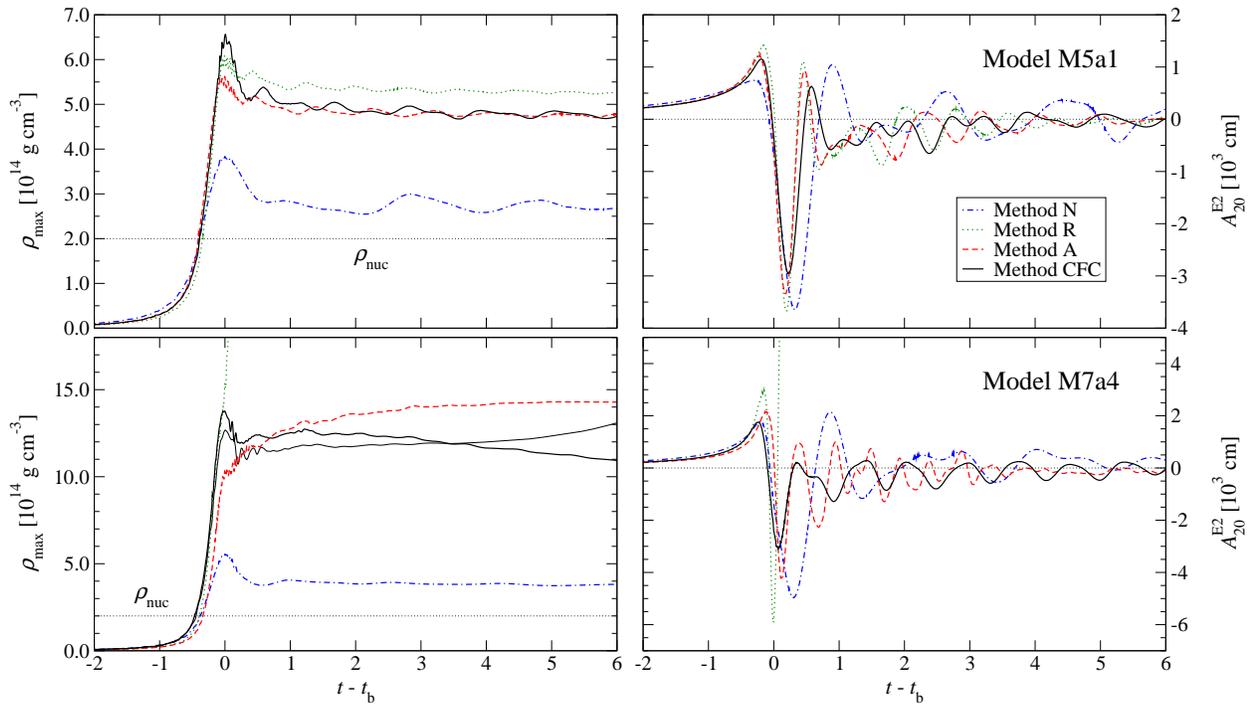}}
  \caption{Time evolution of the maximum density $ \rho_\mathrm{max} $
    (left panels) and the gravitational wave signal amplitude
    $ A^\mathrm{E2}_{20} $ (right panels) for the rotational core
    collapse models~M5a1 (top panels) and~M7a4 (lower panels)
    from simulations using \emph{method~N} (blue dash-dotted lines),
    \emph{method~R} (green dotted lines), \emph{method~A} (red dashed
    lines), and \emph{method~CFC} (black solid lines). The thin black
    solid line in the left bottom panel shows $ \rho_\mathrm{max} $
    for model~M7a4 using \emph{method~CFC} in low resolution.}
  \label{fig:density_and_waveform}
\end{figure}

Fig.~\ref{fig:density_and_waveform} displays the time evolution of the
maximum density $ \rho_\mathrm{\,max} $ and the gravitational wave
signal amplitude $ A^\mathrm{E2}_{20} $ (i.e.\ the gravitational
radiation waveform; for a definition of $ A^\mathrm{E2}_{20} $, see,
e.g., \cite{dimmelmeier_02_b}) for models~M5a1 and~M7a4, which are all
of collapse type NS in a full GR simulation. As the results of
\emph{method~CFC} and~\emph{CFC+} are almost identical, we do not show
the latter in Fig.~\ref{fig:density_and_waveform}. We emphasize that
even for these rather extreme stellar core collapse models, the
2PN deviation from isotropy in \emph{method~CFC+} remains very small,
$ h^\mathrm{\,2PN}_{\,ij} \sim 0.001 \mbox{\,--\,} 0.01 \ll 1 $. Only
for models like M8c4 which reach extremely high densities and low
values of the lapse function, the components of
$ h^\mathrm{\,2PN}_{\,ij} $ increase to 0.02. This further supports
the observation that \emph{method~CFC} already is an excellent
approximation of full GR. For all models the purely Newtonian
simulation, \emph{method~N}, yields maximum densities far below the
values in CFC both during and after core bounce. These results for
rotating \emph{massive} stellar cores are in accordance with the
findings for rotating \emph{regular} stellar cores
in~\cite{dimmelmeier_02_b, dimmelmeier_01_a}. There the typically
higher waveform amplitudes at core bounce in \emph{method~N} compared
to \emph{method~CFC} for many models of type NS, observed here as well
in Fig.~\ref{fig:density_and_waveform}, are also found and explained.

In model~M5a1 (top panel) the post-bounce value of
$ \rho_\mathrm{max} $ for approximation \emph{method~A} matches best
the result with \emph{method~CFC}, which (with \emph{method~CFC+}) is
closest to full GR, while \emph{method~R} gives too large values. A
similar behavior was found and discussed for the rotating core
collapse model~A1B3G1 in~\cite{marek_06_a}. The inadequacy of
\emph{method~R} for reproducing the collapse type of certain models is
evident for model~M7a4 (bottom panel), where the collapse proceeds
unhalted and thus the \emph{prompt} formation of a black hole is
predicted, while the other approximation methods capture the collapse
type right. Beyond employing a good approximation of GR, it is also
crucial to use a numerical grid with sufficiently accuracy. On a
grid with a low central and global resolution of $ n_r = 100 $,
$ n_\theta = 20 $, and
$ \Delta r_\mathrm{\,c} \simeq 500 \mathrm{\ m} $, the core of
model~M7a4 (and many other models as well) simulated in
\emph{method~CFC} suffers from matter fall-back after core bounce,
recollapses, and predicts the \emph{delayed} formation of a black hole
as in~\cite{shibata_05_b} (see left bottom panel of
Fig.~\ref{fig:density_and_waveform}).

%%%%%%%%%%%%%%%%%%%%%%%%%%%%%%%%%%%%%%%%%%%%%%%%%%%%%%%%%%%%%%%%%%%%%%%%%%%%%%%%%%%
%%%%%%%%%%%%%%%%%%%%%%%%%%%%%%%%%%%%%%%%%%%%%%%%%%%%%%%%%%%%%%%%%%%%%%%%%%%%%%%%%%%
%%%%%%%%%%%%%%%%%%%%%%%%%%%%%%%%%%%%%%%%%%%%%%%%%%%%%%%%%%%%%%%%%%%%%%%%%%%%%%%%%%%

\section{Summary and outlook}

We have presented several recently introduced methods for
approximating GR effects in hydrodynamic simulations of stellar core
collapse. When applied to axisymmetric test models for the collapse of
rotating massive stellar cores which reach high densities and rotation
rates, these approximations as expected offer much better results than
a genuinely Newtonian treatment of gravity. While the accuracy of
methods using an effective relativistic potential in an otherwise
Newtonian code degrades at rapid rotation (which can be remedied by a
new formulation~\cite{mueller_06_a}) and these methods reach their
limit for high densities and/or strong rotation, approximations based
on conformal flatness prove to yield excellent matching with full GR
for all investigated models.

%%%%%%%%%%%%%%%%%%%%%%%%%%%%%%%%%%%%%%%%%%%%%%%%%%%%%%%%%%%%%%%%%%%%%%%%%%%%%%%%%%%
%%%%%%%%%%%%%%%%%%%%%%%%%%%%%%%%%%%%%%%%%%%%%%%%%%%%%%%%%%%%%%%%%%%%%%%%%%%%%%%%%%%
%%%%%%%%%%%%%%%%%%%%%%%%%%%%%%%%%%%%%%%%%%%%%%%%%%%%%%%%%%%%%%%%%%%%%%%%%%%%%%%%%%%

\begin{theacknowledgments}
  We gratefully acknowledge contributions from R.~Buras, J.~A.~Font,
  J.~M.~Ib\'a\~nez, H.-T.~Janka, E.~M\"uller, and G.~Sch\"afer to the
  work presented here. This work was supported by the German Research
  Foundation   DFG (SFB/Transregio~7 ``Gravitationswellenastronomie''
  and SFB~375 ``Astroteilchenphysik''), by the EU Network Programme
  (Research Training Network Contract HPRN-CT-2000-00137), and by the
  Spanish Ministerio de Ciencia y Tecnolog\'{\i}a (Grant AYA
  2001-3490-C02-01).
\end{theacknowledgments}

%%%%%%%%%%%%%%%%%%%%%%%%%%%%%%%%%%%%%%%%%%%%%%%%%%%%%%%%%%%%%%%%%%%%%%%%%%%%%%%%%%%
%%%%%%%%%%%%%%%%%%%%%%%%%%%%%%%%%%%%%%%%%%%%%%%%%%%%%%%%%%%%%%%%%%%%%%%%%%%%%%%%%%%
%%%%%%%%%%%%%%%%%%%%%%%%%%%%%%%%%%%%%%%%%%%%%%%%%%%%%%%%%%%%%%%%%%%%%%%%%%%%%%%%%%%

\end{document}